\pdfoutput=1 
\documentclass[array,enumerate,floatfix,latexsym,longbibliography,pra,preprintnumbers,showpacs,twocolumn]{revtex4-1}

\usepackage{amsmath,amssymb}
\usepackage{braket}
\usepackage{graphicx}
\usepackage{hyperref}
\usepackage{centernot}
\usepackage{pgfplots}
\usepackage{relsize}
\usepackage{tikz}
\usetikzlibrary{arrows}
\usetikzlibrary{calc}
\usetikzlibrary{external}

\usetikzlibrary{quantikz}
\usepackage{verbatim}

\newcommand{\Lpp}[1][]{\mathcal{L}^{++}_{ #1}}

\newcommand{\NL}{\mathcal{N}_L}

\newcommand{\NR}{\mathcal{N}_R}
\newcommand{\CNOT}{\textsc{CNOT}}

\begin{document}

\title{Solving Gauss's Law on Digital Quantum Computers\\with Loop-String-Hadron Digitization}

\author{Indrakshi Raychowdhury}
\email{iraychow@umd.edu}
\affiliation{Maryland Center for Fundamental Physics and Department of Physics,
University of Maryland, College Park, MD 20742, USA}

\author{Jesse R.~Stryker}
\email[]{stryker@uw.edu}
\affiliation{Institute for Nuclear Theory, University of Washington, Seattle, WA 98195, USA}

\date{\today}

\preprint{INT-PUB-18-062}

\begin{abstract} 

  We show that using the loop-string-hadron (LSH) formulation of SU(2) lattice gauge theory (arXiv:1912.06133) as a basis for digital quantum computation easily solves an important problem of fundamental interest: implementing gauge invariance (or Gauss's law) exactly.
  We first discuss the structure of the LSH Hilbert space in $d$ spatial dimensions, its truncation, and its digitization with qubits.
  Error detection and mitigation in gauge theory simulations would benefit from physicality ``oracles,'' so we decompose circuits that flag gauge invariant wavefunctions.
  We then analyze the logical qubit costs and entangling gate counts involved with the protocols.
  The LSH basis could save or cost more qubits than a Kogut-Susskind-type representation basis, depending on how that is digitized as well as the spatial dimension.
  The numerous other clear benefits encourage future studies into applying this framework.

\end{abstract}


\maketitle

\section{Introduction}
Quantum computation is expected to efficiently simulate quantum mechanical many-body problems 
\cite{%
  benioffComputerPhysical80,
  feynmanSimulatingPhysics82,
  lloydUniversalQuantum96%
},
as quantum computers can handle the exponential growth of information in entangled quantum systems that overwhelms classical computers.
For lattice gauge theories (LGTs), quantum computers offer hope for \emph{ab initio} studies of non-zero density, topological properties, and real-time phenomena, which are exponentially hard to solve classically due to sign problems.
Rapid progress has been made recently in developing the theory needed to simulate Abelian gauge field theories
\cite{%
  zohar.farace.eaDigitalQuantum17,
  schweizer.grusdt.eaFloquetApproach19,
  rico.pichler.eaTensorNetworks14,
  muschik.heyl.eaWilsonLattice17,
  zache.hebenstreit.eaQuantumSimulation18,
  klco.dumitrescu.eaQuantumclassicalComputation18,
  davoudi.hafezi.eaAnalogQuantum20,
  shaw.lougovski.eaQuantumAlgorithms20%
},
but the leap to their non-Abelian counterparts
\cite{%
  byrnes.yamamotoSimulatingLattice06,
  zohar.cirac.eaColdAtomQuantum13,
  tagliacozzo.celi.eaSimulationNonAbelian13,
  banerjee.bogli.eaAtomicQuantum13,
  wieseUltracoldQuantum13,
  wieseQuantumSimulating14,
  tagliacozzo.celi.eaTensorNetworks14,
  mezzacapo.rico.eaNonAbelianSU15,
  nuqscollaboration.lamm.eaGeneralMethods19,
  nuqscollaboration.alexandru.eaGluonField19,
  raychowdhury.strykerLoopString20,
  klco.savage.eaSUNonAbelian20%
}
continues to look significant.
(For reviews on quantum simulation of gauge theories, see Refs.
\cite{%
  dalmonte.montangeroLatticeGauge16,
  zohar.cirac.eaQuantumSimulations16,
  banuls.cichyReviewNovel20,
  banuls.blatt.eaSimulatingLattice19%
}.)
The most basic issue any simulation proposal must address is how to formulate the gauge theory and map its wavefunctions to those of the quantum computer, but with the limitations faced by quantum hardware in the NISQ era
\cite{preskillQuantumComputing18},
it is crucial that LGTs be formulated economically and tailored to the capabilities of quantum devices
\cite{%
  zohar.cirac.eaColdAtomQuantum13,
  martinez.muschik.eaRealtimeDynamics16,
  klco.savage.eaSUNonAbelian20%
}.

Non-Abelian Hamiltonian LGT is conventionally constructed in terms of a local Hamiltonian and a Hilbert space of group-representation states, supplemented by a non-Abelian Gauss law for gauge invariance
\cite{%
  kogut.susskindHamiltonianFormulation75,
  creutzQuarksGluons84,
  smitIntroductionQuantum02,
  zohar.burrelloFormulationLattice15%
}.
States in the Hilbert space are predominantly unphysical, and a noisy quantum computer will suffer from wandering into the vast space of non-gauge-invariant states.
In order to remove gauge redundancy, one must impose Gauss's law, which is nontrivial on a quantum computer as its color components are not simultaneously diagonalizable.
Apart from that, the different irreducible representations of the gauge group to be mapped onto a register of qubits are on different footings under (and mixed by) the action of the Hamiltonian;
crafting the action of that Hamiltonian in terms of quantum computer operations seems rather unnatural to do.

From the viewpoint of digital quantum simulation, the equivalent loop-string-hadron (LSH) formulation of LGT \cite{raychowdhury.strykerLoopString20} offers some significant advantages as a digitization scheme.
The Hilbert space is characterized in terms of discrete flux and quark excitations, which is to say the states are naturally parametrized in terms of binary fermionic occupation numbers and unbounded bosonic occupation numbers.
In addition, the elementary excitations are intrinsically SU(2)-invariant;
non-Abelian gauge redundancy is therefore completely eradicated.
The problem of unphysical sectors does still persist in the form of Abelian Gauss law constraints requiring flux conservation along links, but the LSH constraints are all simultaneously diagonalized, so basis states are each definitely allowed or definitely unallowed.
The Hamiltonian then mixes states among each other via raising and lowering operations on all the occupation numbers.

In this article, we first apply the loop-string-hadron formulation to digital quantum simulations of non-Abelian gauge theories.
In Sec.\,\,\ref{sec:LSHBasis} we review the elements of the LSH framework for SU(2) gauge fields coupled to one staggered fermion flavor in $d=(1,2,3)$ spatial dimensions  and give a digitization scheme for this basis.
Then in Sec.\,\,\ref{sec:circuits}, we take advantage of the LSH basis to present the first decompositions of SU(2) physicality oracles, i.e., quantum algorithms that probe charge conservation of an SU(2) lattice gauge theory wave function.
Finally, in Sec.\,\,\ref{sec:discussion} we discuss the resources involved with implementation on digital hardware and the known advantages and disadvantages compared to group-representation bases.

\section{\label{sec:LSHBasis}Loop-String-Hadron basis}

The LSH formulation \cite{raychowdhury.strykerLoopString20} of SU(2) Hamiltonian lattice gauge theory is a derivative of the so-called prepotential formulation \cite{
  mathurHarmonicOscillator05,
  mathurLoopStates06,
  mathurLoopApproach07,
  anishetty.mathur.eaPrepotentialFormulation10,
  raychowdhuryPrepotentialFormulation13%
},
combined with matter fields, and is equivalent to the more conventional Kogut-Susskind formulation in which the degrees of freedom are electric fields and link variables from the gauge group.
The basic objects in the LSH framework are a collection of operators associated with the sites of the lattice:
local ``loop,'' ``string,'' and ``hadron'' operators, named for the types of excitations they manipulate at a site.
The corresponding operator algebra and derivation of the Hamiltonian in terms of these objects has been worked out in detail in \cite{raychowdhury.strykerLoopString20}.
For the purposes of this work we are primarily interested in the Hilbert space structure derived from these degrees of freedom.

\subsection*{Lattice geometry and topology}

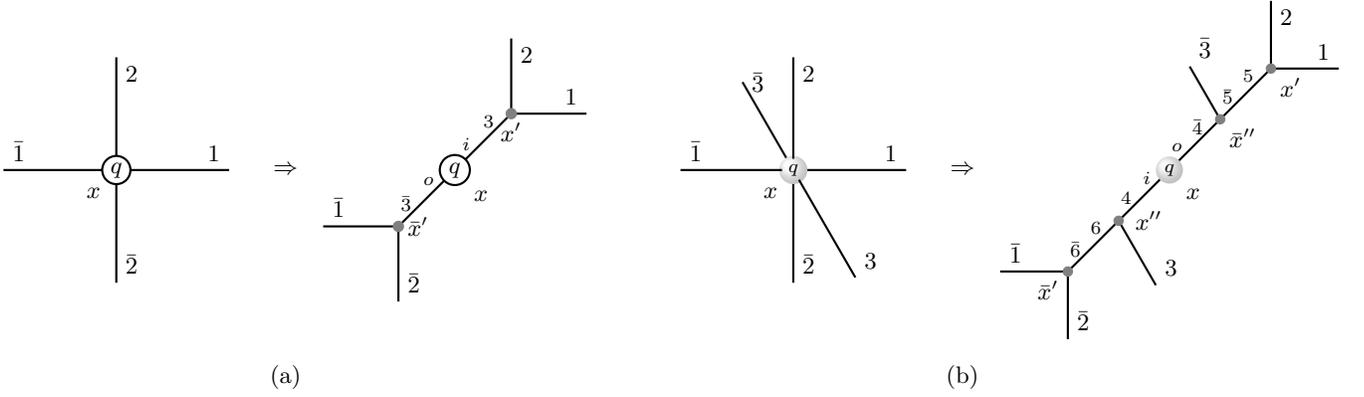
\begin{figure*}
\begin{center}
  \begin{tikzpicture}
    \coordinate (A) at ( 0, 0);
    \coordinate (2dOrigin) at ( 2.25, 0);
    \coordinate (C) at ( 4.5, 0);

    \coordinate (E) at ( 9, 0);
    \coordinate (3dOrigin) at ( 11.25, 0);
    \coordinate (G) at ( 14, 0);

    \node at (A) {
      \begin{tikzpicture}[scale=1.5, thick]
      \draw (-1,0) node[above right] {$\bar{1}$} -- (1,0) node[above left] {$1$};
      \draw (0,-1) node[above right] {$\bar{2}$} -- (0,1) node[below right] {$2$};
      \filldraw [fill=white, draw=black] (0,0) circle [radius=0.125] node {$q$};
      \draw (-0.2,-0.2) node {$x$};
        \end{tikzpicture}
    };

    \draw (2dOrigin) node {$\quad\Rightarrow\quad$};
    \draw ($(2dOrigin) + ( 0, -2.75)$) node {(a)};

    \node at (C) {
  \begin{tikzpicture}[scale=1., thick]
      \draw (-1,0) node[above right] {$\bar{1}$} -- (0,0);
      \draw (0,-1) node[above right] {$\bar{2}$} -- (0,0);
      \draw (1.5,1.5) -- (2.5,1.5) node[above left] {$1$};
      \draw (1.5,1.5) -- (1.5,2.5) node[below right] {$2$};
      \draw (0,0) node[right] {$\bar{x}^{\prime}$} node [shift={(0.1,0.3)}] {\scriptsize $\bar{3}$}
      -- (1.5,1.5) node[below] {$x^{\prime}$} node [shift={(-0.3,-0.1)}] {\scriptsize $3$};
      \filldraw [fill=white, draw=black] (0.75,0.75) circle [radius=0.2]
      node {$q$} node [shift={(-0.35,-0.15)}] {\scriptsize $o$} node [shift={(0.15,0.35)}] {\scriptsize $i$};
      \draw (1.1,0.4) node {$x$};
      \filldraw [fill=gray, draw=gray] (0,0) circle [radius=0.06125];
      \filldraw [fill=gray, draw=gray] (1.5,1.5) circle [radius=0.06125];
  \end{tikzpicture}
      };

    \draw (3dOrigin) node {$\quad\Rightarrow\quad$};
    \draw ($(3dOrigin) + ( 0, -2.75)$) node {(b)};

    \node at (E) {
  \begin{tikzpicture}[scale=1.5, thick]
    \begin{scope}
      \draw (0,0) -- (1,0) node[above left] {$1$};
      \draw (0,-1) node[above right] {$\bar{2}$} -- (0,0);
      \draw (0,0) -- ({0.9 * cos(120)}, {0.9 * sin(120)}) node[right] {$\bar{3}$};
      \fill [fill=white] (0,0) circle [radius=0.125];
      \shade[ball color = gray!15, opacity = 0.4] (0,0) circle (.125);
      \draw (0,0) node {\scriptsize{$q$}};
      \draw (-1,0) node[above right] {$\bar{1}$} -- (-0.1,0);
      \draw (0,0.1) -- (0,1) node[below right] {$2$};
      \draw ({1.1 * cos(300)}, {1.1 * sin(300)})  node[above right] {$3$} -- ({0.1 * cos(300)}, {0.1 * sin(300)});
      \draw (-0.2,-0.2) node {$x$};
    \end{scope}
  \end{tikzpicture}
    };

    \node at (G) {
  \begin{tikzpicture}[scale=0.9, thick]
    \begin{scope}
      \coordinate (A) at (1.5, 1.5);
      \coordinate (B) at (0.75, 0.75);
      \coordinate (C) at (-0.75, -0.75);
      \coordinate (D) at (-1.5, -1.5);
      \draw (A) -- ($(A)+(1,0)$) node[above left] {$1$};
      \draw (A) -- ($(A)+(0,1)$) node[below right] {$2$};
      \draw ($(D)+(-1,0)$) node[above right] {$\bar{1}$} -- (D);
      \draw ($(D)+(0,-1)$) node[above right] {$\bar{2}$} -- (D);
      \draw (D) node [shift={(0.1,0.3)}] {\scriptsize $\bar{6}$}
      -- (C) node [shift={(-0.3,-0.1)}] {\scriptsize $6$};
      \draw (C) node [shift={(0.1,0.3)}] {\scriptsize $4$}
      -- (0,0) node [shift={(-0.3,-0.1)}] {\scriptsize $i$};
      \draw (0,0) node [shift={(0.1,0.3)}] {\scriptsize $o$}
      -- (B) node [shift={(-0.3,-0.1)}] {\scriptsize $\bar{4}$};
      \draw (B) node [shift={(0.1,0.3)}] {\scriptsize $\bar{5}$}
      -- (A) node [shift={(-0.3,-0.1)}] {\scriptsize $5$};
      \draw ($(C)+({1.1 * cos(300)}, {1.1 * sin(300)})$) node[above right] {$3$} -- (C);
      \draw ($(B)+({0.9 * cos(120)}, {0.9 * sin(120)})$) node[above right] {$\bar{3}$} -- (B);
      \fill [fill=white] (0, 0) circle [radius=0.2];
      \shade[ball color = gray!25, opacity = 0.4] (0,0) circle (0.2);
      \draw (0,0) node {\scriptsize{$q$}};
      \draw (0.35,-0.35) node {$x$};
      \filldraw [fill=gray, draw=gray] (D) circle [radius=0.06125] node[below left] {$\bar{x}^{\prime}$};
      \filldraw [fill=gray, draw=gray] (C) circle [radius=0.06125] node [shift={(0.1,0)}, right] {$x^{\prime\prime}$};
      \filldraw [fill=gray, draw=gray] (B) circle [radius=0.06125] node [below right] {$\bar{x}^{\prime\prime}$};
      \filldraw [fill=gray, draw=gray] (A) circle [radius=0.06125] node [below right] {$x^{\prime}$};
    \end{scope}
  \end{tikzpicture}
    };
  \end{tikzpicture}
\end{center}
\caption{
\label{fig:lshSites}
  (a) Virtual point splitting of a 2D lattice site $x$ into gluonic sites $x', \bar{x}'$, with matter living on the central quark site $x$.
  (b) Virtual point splitting of a 3D lattice site into gluonic sites $x, x', \bar{x}', x'', \bar{x}''$ connected by internal links, with matter again at the central quark site $x$.
}
\end{figure*}

The full suite of LSH operators available depends on the topology of the lattice.
We consider Cartesian lattices in $d$ spatial dimensions.
The $2d$-point vertices of the lattice are formally split into a network of three-point vertices by introducing virtual (internal) links and sites (Fig.~\ref{fig:lshSites}), whose role is to avoid an overcomplete basis.
The three-point ``gluonic vertices'' serve as junctions for gauge flux to pass through.
Quarks are accommodated along virtual links by further dividing one of them in half, with ``quark vertices'' residing at virtual two-point vertices connecting the halves.
(The virtual degrees of freedom are bookkeeping devices that do not count toward the electric energy and the resulting structure encodes equivalent dynamics to the Kogut-Susskind formulation \cite{raychowdhuryLowEnergy19,anishetty.sreerajMassGap18}.)

In one-dimensional (1D) space only, the virtual ``point splitting'' procedure described is not necessary and all sites of the lattice are quark sites.

The virtual splitting of a 2D lattice site shown in Fig. \ref{fig:lshSites}(a) results topologically in a hexagonal lattice.
The matter field is situated in the middle of the virtual link, dividing the $3-\bar{3}$ link into two.
The gluonic vertices are referred to as $x'$ and $x''$, while the quark vertex is simply referred to by the lattice site coordinate $x$.

In three-dimensional (3D) space, each site splits into four three-point vertices and matter is added at another intermediate two-point vertex, as shown in Fig. \ref{fig:lshSites}(b).
In Fig.\,\,\ref{fig:lshSites}(b), the 4-$\bar{4}$ link was divided to host the quark site.

\subsection*{Local structures}

Regardless of the spatial dimension, the virtual topology is such that there are only two possible elementary structures---the quark sites and the gluonic sites.
We now review these structures.
\begin{figure}
  \begin{center}
  \begin{tikzpicture}[scale=0.75]
    \begin{scope}[thick]
      \draw ( -8, 0) -- ( -4, 0);
      \filldraw [fill=white, draw=black] ( -6, 0) circle [radius=0.0625]
      node [xshift=0, yshift=-10] {$x$}
      node [xshift=-15, yshift=-6] {\footnotesize $i$}
      node [xshift=15, yshift=-6] {\footnotesize $o$};
      \draw ( 0, 0) -- ( 2, 0) node [below] {$p$};
      \draw ( 0, 0) -- ( {2 * cos(120)}, {2 * sin(120)} ) node [below left] {$q$};
      \draw ( 0, 0) -- ( {2 * cos(240)}, {2 * sin(240)} ) node [above left] {$r$};
      \draw ( 0, 0) node [below right] {$x_g$};
    \end{scope}
  \end{tikzpicture}
    \caption{\label{fig:virtualSites}
      Depictions of the virtual sites belonging to the point-split lattice.
      Left: A quark site.
      Right: A gluonic site.
    }
  \end{center}
\end{figure}
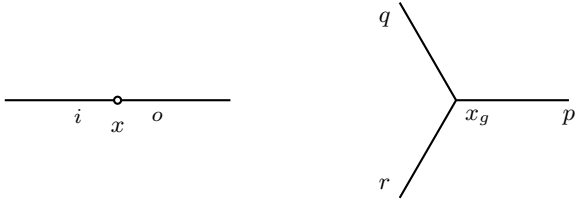

\begin{center}
  \textsc{Quark sites}
\end{center}

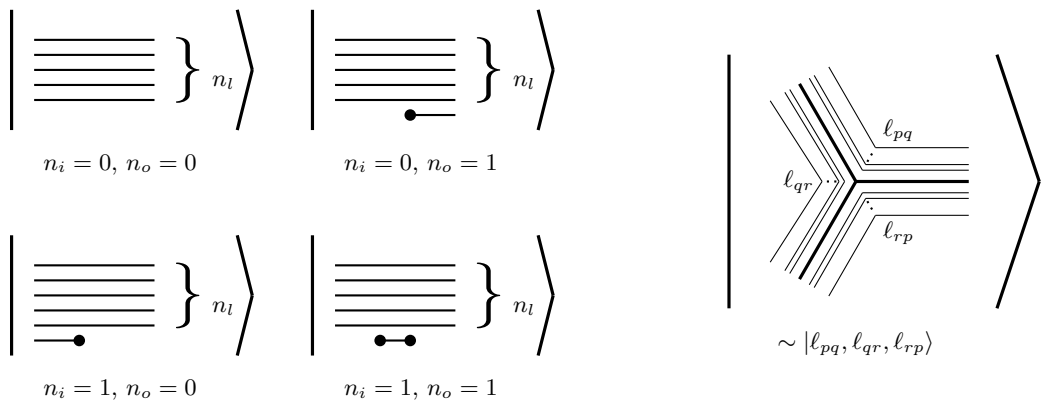
\begin{figure*}
  \begin{center}
  \begin{tikzpicture}
    \coordinate (quarkO) at ( 0, 0);
    \coordinate (gluonO) at ( 8, 0);

    \node at (quarkO) {
      \begin{tikzpicture}[scale=2.0, thick]
    \begin{scope}
      \coordinate (A) at ( -1, 0.75);
      \coordinate (B) at ( 1, 0.75);
      \coordinate (C) at ( -1, -0.75);
      \coordinate (D) at ( 1, -0.75);

      \foreach \X/\y in { -1/0.75, 1/0.75, -1/-0.75, 1/-0.75} { 
        \draw[very thick] ($( \X, \y) + ( -0.15, -0.4)$) -- ($( \X, \y) + ( -0.15, 0.4)$);
        \draw[very thick] ($( \X, \y) + ( 1.45, 0)$) -- ($( \X, \y) + ( 1.35, +0.4)$);
        \draw[very thick] ($( \X, \y) + ( 1.45, 0)$) -- ($( \X, \y) + ( 1.35, -0.4)$);
        \foreach \dy in { -0.2, -0.1, 0, 0.1, 0.2}
          \draw ($( \X, \y) + ( 0, \dy)$) -- ($( \X, \y) + ( 0.8, \dy)$);
        \draw ($( \X, \y) + ( 0.85, 0)$) node [right] {{\Huge $\rbrace$} $n_l$ };
      };

      \draw ($(B) + ( +0.5, -0.3)$) -- ($(B) + ( +0.8, -0.3)$);
      \filldraw [fill=black, draw=black] ($(B) + ( +0.5, -0.3)$) circle [radius=0.03125];
      \draw ($(C) + ( +0.3, -0.3)$) -- ($(C) + ( 0, -0.3)$);
      \filldraw [fill=black, draw=black] ($(C) + ( +0.3, -0.3)$) circle [radius=0.03125];
      \draw ($(D) + ( +0.3, -0.3)$) -- ($(D) + ( +0.5, -0.3)$);
      \filldraw [fill=black, draw=black] ($(D) + ( +0.3, -0.3)$) circle [radius=0.03125];
      \filldraw [fill=black, draw=black] ($(D) + ( +0.5, -0.3)$) circle [radius=0.03125];
      \draw ($(A) + ( +0.0, -0.5)$) node[below right] {$n_i=0$, $n_o=0$};
      \draw ($(B) + ( +0.0, -0.5)$) node[below right] {$n_i=0$, $n_o=1$};
      \draw ($(C) + ( +0.0, -0.5)$) node[below right] {$n_i=1$, $n_o=0$};
      \draw ($(D) + ( +0.0, -0.5)$) node[below right] {$n_i=1$, $n_o=1$};
    \end{scope}
  \end{tikzpicture}
    };

    \node at (gluonO) {
  \begin{tikzpicture}
    \begin{scope}[scale=0.75]
      \coordinate (O) at ( 0, 0);
      \coordinate (P) at ( 2, 0);
      \coordinate (Q) at ( {2 * cos(120)}, {2 * sin(120)} );
      \coordinate (R) at ( {2 * cos(240)}, {2 * sin(240)} );
      \draw[very thick] ( 0, 0) -- ( 2, 0);
      \draw[very thick] ( 0, 0) -- ( {2 * cos(120)}, {2 * sin(120)} );
      \draw[very thick] ( 0, 0) -- ( {2 * cos(240)}, {2 * sin(240)} );
      \foreach \d in {0.2,0.3,0.6}{
        \draw ($(O) + ( {\d * cot(60)}, {\d * -1} )$) -- ($(P) + ( 0, -\d)$);
        \draw ($(O) + ( {\d * cot(60)}, {\d * 1} )$) -- ($(P) + ( 0, \d)$);
        \draw ($(O) + ( {\d * cot(60)}, {\d * 1} )$) -- ($(Q) + ( {\d * sin(60)}, {\d * cos(60)})$);
        \draw ($(O) + ( {\d * -1}, 0 )$) -- ($(Q) + ( {\d * -sin(60)}, {\d * -cos(60)})$);
        \draw ($(O) + ( {\d * -1}, 0 )$) -- ($(R) + ( {\d * -sin(60)}, {\d * cos(60)})$);
        \draw ($(O) + ( {\d * cot(60)}, {\d * -1} )$) -- ($(R) + ( {\d * sin(60)}, {\d * -cos(60)})$);
      };
      \draw[thick, dotted] ($(O) + ( {0.35 * cot(60)}, {0.35 * -1} )$) -- ($(O) + ( {0.55 * cot(60)}, {0.55 * -1} )$) node [below right] {$\ell_{rp}$};
      \draw[thick, dotted] ($(O) + ( {0.35 * cot(60)}, {0.35 * 1} )$) -- ($(O) + ( {0.55 * cot(60)}, {0.55 * 1} )$) node [above right] {$\ell_{pq}$};
      \draw[thick, dotted] ($(O) + ( {-0.3 / sin(60)}, 0 )$) -- ($(O) + ( {-0.5 / sin(60)}, 0 )$) node [left] {$\ell_{qr}$};
      \draw[very thick] ($(O) + ( -2.25, -2.25)$) -- ($(O) + ( -2.25, 2.25)$);
      \draw[very thick] ($(O) + ( 3.25, 0)$) -- ($(O) + ( 2.5, -2.25)$);
      \draw[very thick] ($(O) + ( 3.25, 0)$) -- ($(O) + ( 2.5, +2.25)$);
      \draw ($(O) + ( 0, -2.5)$) node [below] {$\sim \ket{\ell_{pq}, \ell_{qr}, \ell_{rp}}$};
    \end{scope}
  \end{tikzpicture}
    };
  \end{tikzpicture}
    \end{center}
    \caption{\label{fig:siteKets}
      Left: Depiction of SU(2)-invariant configurations at a quark site.
      The on-site state is characterized by a loop quantum number $n_l$ (the solid lines) and two quark quantum numbers $n_i$ and $n_o$ (the blobs).
      Right: Depiction of SU(2)-invariant configurations at a gluonic site.
      The on-site state is characterized by three loop quantum numbers $\ell_{pq}$, $\ell_{qr}$, and $\ell_{rp}$, with $\ell_{ij}$ counting the flux lines flowing into $i$ and out of $j$.
    }
\end{figure*}

At a quark site (left panel of Fig.~\ref{fig:virtualSites}), the collection of operators [for SU(2) gauge fields coupled to one staggered quark] consists of the loop ($\mathcal{L}$), string ($\mathcal{S}_{\text{in}/\text{out}}$), and hadron ($\mathcal{H}$) operators, along with number operators ($\mathcal{N}$) that count quark number or flux strength on either side of the site.
All possible site-local SU(2)-singlet excitations generated by these operators can be characterized in terms of one bosonic loop quantum number $n_l$, and two fermionic occupancies $n_i$ and $n_o$ (for ``in'' and ``out'').
We denote these by
\begin{align}
  \ket{n_l, n_i, n_o}: & \nonumber \\
  \label{eq:LSH1dQuantumNumbersRange} n_l & = 0,1,2,\cdots \quad n_i  =0,1 \quad n_o=0,1 \ .
\end{align}
The loop quantum number $n_l$ measures how many units of pure gauge flux pass through the site.
An important distinction of SU(2) gauge flux relative to a U(1) gauge theory is that it lacks directionality;
whether a flux unit actually comes ``in'' from one side and goes ``out'' the other or vice versa is only a manner of speaking.
The occupancy numbers $n_i$ and $n_o$ indicate quark content---exactly one of these equaling 1 implies it is part of a string (joined to a unit of flux on one side), while both equaling 1 signals they are paired up into a hadron.
(Ensuring anticommuting statistics is a dynamical issue irrelevant to this paper.)
Visual representations of local quark site states are displayed in the left panel of Fig. \ref{fig:siteKets}.

\begin{center}
  \textsc{Gluonic sites}
\end{center}

A point-split lattice site has $2(d-1)$ distinct gluonic sites.
At a gluonic site  $x_g$ (right panel of Fig.~\ref{fig:virtualSites}), the three attached links can be labeled with integers $p,q,r$ such that $p<q<r$.
In 2D space, the labels are simply $(p,q,r)=(1,2,3)$ or $(p,q,r)=(\bar{1},\bar{2},\bar{3})$ but a variety of $(p,q,r)$ combinations are involved in $d>2$.

The local Hilbert space at $x_g$ is generated by loop creation operators ($\Lpp[pq]$, $\Lpp[qr]$, $\Lpp[rp]$).
All possible SU(2)-singlet states can therefore be constructed from elementary flux units that individually link two distinct directions.
The intuitive illustration of these states is seen in the right panel of Fig.~\ref{fig:siteKets}.
A basis for a local gluonic-site Hilbert space is denoted by
\begin{align}
  \ket{\ell_{pq},\ell_{qr},\ell_{rp}}: & \nonumber \\
  \label{eq:LSHGluonicQuantumNumbersRange} \ell_{ij} & = 0, 1, 2, \cdots (ij=pq,qr,rp).
\end{align}

\subsection*{Gauge constraints}

The last major ingredient of the Hilbert space is the Abelian Gauss law constraints.
The Abelian Gauss law is the same physical requirement across all links of the lattice, in any dimension:
the total flux magnitudes at each end of a link must be equal.

A given link end within the point-split lattice attaches to either a quark site or a gluonic site.
If that link end attaches to a quark site, the relevant flux number operator is denoted by $\NL$ ($\NR$) if the link end is ``outgoing'' (``incoming'') relative to the site.
In particular,
\begin{subequations}
\begin{align}
  \NL \ket{n_l,n_i,n_o} &= [n_l + n_o (1-n_i)]\ket{n_l,n_i,n_o}\\
  \NR \ket{n_l,n_i,n_o} &= [n_l + n_i (1-n_o)]\ket{n_l,n_i,n_o}
\end{align}
\end{subequations}
If instead the link end attaches to a gluonic site, say, along direction $i(=p,q,r)$ relative to the site, the relevant flux number operator is denoted by $\mathcal{N}_i$.
In this case,
\begin{subequations}
  \label{eq:gluonicFluxNumbers}
\begin{align}
  \mathcal{N}_p \ket{\ell_{pq},\ell_{qr},\ell_{rp}} &= (\ell_{rp} + \ell_{pq}) \ket{\ell_{pq},\ell_{qr},\ell_{rp}} \\
  \mathcal{N}_q \ket{\ell_{pq},\ell_{qr},\ell_{rp}} &= (\ell_{pq} + \ell_{qr}) \ket{\ell_{pq},\ell_{qr},\ell_{rp}} \\
  \mathcal{N}_r \ket{\ell_{pq},\ell_{qr},\ell_{rp}} &= (\ell_{qr} + \ell_{rp}) \ket{\ell_{pq},\ell_{qr},\ell_{rp}}
\end{align}
\end{subequations}
Using these, the Abelian Gauss law is the requirement that, for any link of the lattice, the number operators at each end agree.
(This corresponds to the equality of left and right Casimirs that is automatic in Kogut-Susskind formulations.)
For brevity we refer to links joining two gluonic sites as `$gg$' links, a quark site and a gluon site as `$qg$' links, and two quark sites (for $d=1$) as `$qq$' links.

In one spatial dimension every link joins two quark sites, and the Abelian Gauss laws read
\begin{equation}
  \left[ \NL(x) - \NR(x+1) \right] \ket{\mathrm{phys}} = 0
\end{equation}
for every site $x$.
Meanwhile, the 2D Abelian Gauss laws along the three directions of the hexagonal lattice are
\begin{subequations}
\begin{align}
  [ \mathcal{N}_j(x^{\prime}) - \mathcal{N}_{\bar j}(\overline{x+e_j}^{\,\prime}) ] \ket{\mathrm{phys}} &=  0  \qquad (j=1,2) \ , \\
  [ \mathcal{N}_3 (x^\prime) - \mathcal{N}_R (x) ] \ket{\mathrm{phys}} &= 0 \ , \\
  [ \mathcal{N}_{\bar{3}} (\bar{x}^{\prime}) -  \mathcal{N}_L (x) ] \ket{\mathrm{phys}} &= 0 .
\end{align}
\end{subequations}
And in 3D space, using the labeling from Fig.~\ref{fig:lshSites}(b), the Abelian Gauss laws read
\begin{subequations}
\begin{align}
  [ \mathcal{N}_i(x^{\prime}) - \mathcal{N}_{\bar i}(\overline{x+e_i}^{\,\prime}) ] \ket{\mathrm{phys}} &= 0 \ , \quad (i=1,2) \\
  [ \mathcal{N}_3(x '') - \mathcal{N}_{\bar{3}}(\overline{x+e_3}'') ] \ket{\mathrm{phys}} &= 0 \ , \\
  [ \mathcal{N}_5(x ') - \mathcal{N}_{\bar{5}}(\bar{x}'') ] \ket{\mathrm{phys}} &= 0 \ , \\
  [ \mathcal{N}_6(x '') - \mathcal{N}_{\bar{6}}(\bar{x}') ] \ket{\mathrm{phys}} &= 0 \ , \\
  [ \mathcal{N}_4(x '') - \mathcal{N}_i(x) (1-\mathcal{N}_o(x)) ] \ket{\mathrm{phys}} &= 0 \ , \\
  [ \mathcal{N}_{\bar{4}}(\bar{x} '') - \mathcal{N}_o(x) (1-\mathcal{N}_i(x)) ] \ket{\mathrm{phys}} &= 0 \ .
\end{align}
\end{subequations}

The Abelian constraints above stand in contrast to the ordinary situation in Hamiltonian lattice gauge theory.
The chosen degrees of freedom do not automate flux conservation along links, which is why the constraint is Abelian flux conservation.
Ordinarily flux conservation along links is automatic, and color charge conservation must be imposed at vertices.

Naively, checking non-Abelian gauge invariance would pose a significant practical challenge on a quantum computer.
SU(2) Gauss law operators take the form
\begin{align}
  G^{\mathrm{a}}(x) = \sum_{i=1}^{d} \left[ J^{\mathrm{a}}_{L,i} (x) + J^{\mathrm{a}} _{R,i} (x) \right] + q^\dagger(x)  (\sigma^{\mathrm a}/2) q (x)
\end{align}
where $\mathrm{a}$=(1,2,3) is a color index, $i$ is a spatial index, $J^{\mathrm{a}}_{L/R,i}$ are the left and right electric field generators, and $q$ is a two-component staggered fermion field for this example.
To confirm $G^{\mathrm a} (x) G^{\mathrm a} (x) =0$, it would be sufficient to confirm two of the color components annihilate a given wavefunction.
The most conventional basis choice would be one that diagonalizes the 3-components, i.e., all the $J^{\mathrm{3}}_i$ and $\rho^3$, and the Casimirs $\sum_{\mathrm a} J^{\mathrm{a}}_i J^{\mathrm{a}}_i$.
One could then directly evaluate $G^3(x)$ using diagonal operations, like in a U(1) gauge theory \cite{strykerOraclesGauss19}.
To check another component, say $G^1(x)$, one could simulate an internal color rotation about the 2-axis on every electric link register $\ket{j,m}$ and on the fermion doublet---requiring $2d+1$ basis transformations.
Such rotations would have to isolate every $j$ sector and mix them in just the right way.
At that point $G^1(x)$ could be checked like $G^3(x)$ was, and the bases could be rotated back.

Better protocols than the above could exist, but digital quantum algorithms have yet to be furnished to validate states subjected to non-Abelian constraints.
This technical endeavor would likely suffer from sensitivity to the chosen digitization.
We instead pursue the problem in the LSH basis with Abelian constraints.

A convenient feature of the Abelian Gauss law constraints, from the viewpoint of quantum simulation, is that they all commute.
The LSH framework simultaneously diagonalizes these constraints so that basis states are definitely allowed or definitely unallowed.
In contrast, in the conventional Kogut-Susskind framework, the only basis state that is gauge invariant is that with vanishing electric fields.

By the same token, since the LSH basis simultaneously diagonalizes all the constraints, projective measurements in this basis do not spoil gauge invariance.
With basis states mapped to the computational basis of a quantum computer, (noiseless) readout of the qubit registers will therefore conserve charge.
This could also have benefits for preparing entangled states that satisfy gauge constraints.

\subsection*{Truncation and binary representation}

To represent the theory on a quantum computer we first truncate the Hilbert space on irreducible representations of the gauge group.
The representations are labeled by angular momenta $j$, related to the flux number operators by $j=\mathcal{N}/2$.
To simulate all states with SU(2) representations up to and including spin $\bar{j}$, it is sufficient to use the following qubit resources:
\begin{enumerate}
  \item $N+1$ qubits per quark site loop number $n_\ell$,
  \item $N$ qubits per gluonic site loop number $\ell_{ij}$,
\end{enumerate}
where
\begin{equation}
  \label{eq:lshTruncation} N = \lceil \log_2 ( \bar{j} + 1 ) \rceil \ .
\end{equation}
The quark occupancy numbers require no truncation.

The loop quantum numbers can then be represented by a computational basis of binary integers.
For example, with the loop numbers expressed in binary form by
\begin{align}
  n_{\ell} &= \sum_{m=0}^{N} 2^m n_{\ell,m} \ \qquad (n_{\ell,m}=0,1) \ , \\
  \ell_{ij} &= \sum_{m=0}^{N-1} 2^m \ell_{ij,m} \ \qquad (\ell_{ij,m}=0,1) \ ,
\end{align}
their associated kets are
\begin{align}
  \ket{n_{\ell}} &= \bigotimes_{m=0}^{N} \ket{n_{\ell,m}} \ , \\
  \ket{\ell_{ij}} &= \bigotimes_{m=0}^{N-1} \ket{\ell_{ij,m}} \ .
\end{align}
Here and below the quantum computing-related conventions follow Ref.\,\,\cite{nielsen.chuangQuantumComputation11}.

Note how straightforward it is to truncate the above mapping -- we simply choose a maximum flux saturation strength and require links to have matching flux on either side.
Were we to keep the color-projection quantum numbers ($J_i^3$), a choice would have to be made regarding how to represent the $\sum_{n=1}^{2\bar{j}+1} n^2$ states on a qubit register and how to exclude some ``extra'' states in the register from dynamics.
That is all in addition to having to satisfy the non-Abelian constraints within the mapped representations.

The lattice Hilbert space can be regarded as a tensor product space of all the site-local Hilbert spaces (provided a simulation accounts for the fermionic nature of the quark numbers).
In practice a volume truncation and choice of boundary conditions are implicit, but we are concerned with only the local structure in the bulk.

\begin{figure*}[t]
  \begin{center}
 \begin{quantikz}[row  sep=0.35cm]  
 \lstick{$  \ket{  n_o  }_{x+1}  $} &\octrl{1}&\qw&\qw&\qw&\qw&\qw&\qw&\qw&\qw&\qw&\octrl{1}&\qw& \rstick{$  \ket{  n_o  }_{x+1}  $} & \\  %
 \lstick{$  \ket{  n_i  }_{x+1}  $} &\ctrl{1}&\qw&\qw&\qw&\qw&\qw&\qw&\qw&\qw&\qw&\ctrl{1}&\qw& \rstick{$  \ket{  n_i  }_{x+1}  $} & \\  %
 \lstick{$  \ket{  0  }  $} &\targ{}&\gate[2]{A'}&\qw&\qw&\qw&\qw&\qw&\qw&\qw&\gate[2]{A'}&\targ{}&\qw& \rstick{$  \ket{  0  }  $} & \\  %
 \lstick{$  \ket{  0  }  \ket{  n_{\ell}  }_{x+1}  $} & \qw  \qwbundle{N+2} && \qw  \qwbundle{N+2} & \gate[5,nwires={2,3,4},disable  auto  height]{\begin{tikzpicture}[scale=0.6]  \coordinate (A) at ( 0.25, -0.4); \coordinate (B) at ( 0.5, -0.8); \coordinate (C) at ( 0.75, -1.2); \coordinate (D) at ( 1.0, -1.6); \foreach \dy in { -0.4, -0.8, -1.6}{   \draw ($( 0, 3) + ( 0, \dy)$) -- ($ ( 1.25, 3) + ( 0, \dy)$);   \draw ($( 0, 0) + ( 0, \dy)$) -- ($ ( 1.25, 0) + ( 0, \dy)$); }; \filldraw [fill=black, draw=black] ($ (A) + ( 0, 3) $)  circle [radius=0.0625]; \filldraw [fill=black, draw=black] ($ (B) + ( 0, 3) $)  circle [radius=0.0625]; \filldraw [fill=black, draw=black] ($ (D) + ( 0, 3) $)  circle [radius=0.0625]; \draw (A) circle [radius=0.125]; \draw ($(A) + ( -0.125, 0)$) -- ($(A) + ( +0.125, 0)$); \draw ($(A) + ( 0, -0.125)$) -- ($(A) + ( 0, +0.125)$); \draw (B) circle [radius=0.125]; \draw ($(B) + ( -0.125, 0)$) -- ($(B) + ( +0.125, 0)$); \draw ($(B) + ( 0, -0.125)$) -- ($(B) + ( 0, +0.125)$); \draw (D) circle [radius=0.125]; \draw ($(D) + ( -0.125, 0)$) -- ($(D) + ( +0.125, 0)$); \draw ($(D) + ( 0, -0.125)$) -- ($(D) + ( 0, +0.125)$); \foreach \y in { 3, 0} {   \draw ($ (C) + ( 0, \y) $) node {$\cdot$};   \draw ($ (C) + ( 0, \y) + ( -0.0625, 0.1) $) node {$\cdot$};   \draw ($ (C) + ( 0, \y)  + ( 0.0625, -0.1)$) node {$\cdot$}; }; \draw (A) -- ($ (A) + ( 0, 3) $); \draw (B) -- ($ (B) + ( 0, 3) $); \draw (D) -- ($ (D) + ( 0, 3) $);  \end{tikzpicture}  }   &\qw& \qw  \qwbundle{N+2} &\qw& \gate[5,nwires={2,3,4},disable  auto  height]{\begin{tikzpicture}[scale=0.6]  \coordinate (A) at ( 0.25, -0.4); \coordinate (B) at ( 0.5, -0.8); \coordinate (C) at ( 0.75, -1.2); \coordinate (D) at ( 1.0, -1.6); \foreach \dy in { -0.4, -0.8, -1.6}{   \draw ($( 0, 3) + ( 0, \dy)$) -- ($ ( 1.25, 3) + ( 0, \dy)$);   \draw ($( 0, 0) + ( 0, \dy)$) -- ($ ( 1.25, 0) + ( 0, \dy)$); }; \filldraw [fill=black, draw=black] ($ (A) + ( 0, 3) $)  circle [radius=0.0625]; \filldraw [fill=black, draw=black] ($ (B) + ( 0, 3) $)  circle [radius=0.0625]; \filldraw [fill=black, draw=black] ($ (D) + ( 0, 3) $)  circle [radius=0.0625]; \draw (A) circle [radius=0.125]; \draw ($(A) + ( -0.125, 0)$) -- ($(A) + ( +0.125, 0)$); \draw ($(A) + ( 0, -0.125)$) -- ($(A) + ( 0, +0.125)$); \draw (B) circle [radius=0.125]; \draw ($(B) + ( -0.125, 0)$) -- ($(B) + ( +0.125, 0)$); \draw ($(B) + ( 0, -0.125)$) -- ($(B) + ( 0, +0.125)$); \draw (D) circle [radius=0.125]; \draw ($(D) + ( -0.125, 0)$) -- ($(D) + ( +0.125, 0)$); \draw ($(D) + ( 0, -0.125)$) -- ($(D) + ( 0, +0.125)$); \foreach \y in { 3, 0} {   \draw ($ (C) + ( 0, \y) $) node {$\cdot$};   \draw ($ (C) + ( 0, \y) + ( -0.0625, 0.1) $) node {$\cdot$};   \draw ($ (C) + ( 0, \y)  + ( 0.0625, -0.1)$) node {$\cdot$}; }; \draw (A) -- ($ (A) + ( 0, 3) $); \draw (B) -- ($ (B) + ( 0, 3) $); \draw (D) -- ($ (D) + ( 0, 3) $);  \end{tikzpicture}  }   & \qw  \qwbundle{N+2} &&\qw& \qw  \qwbundle{N+2} & \rstick{$  \ket{  0  }  \ket{  n_{\ell}  }_{x+1}  $} & \\  %
 \lstick{$  \ket{  n_o  }_{x}  $} &\ctrl{1}&\qw&\qw&&&\qw&\qw&&&\qw&\ctrl{1}&\qw& \rstick{$  \ket{  n_o  }_{x}  $} & \\  %
 \lstick{$  \ket{  n_i  }_{x}  $} &\octrl{1}&\qw&\qw&&&\qw&\qw&&&\qw&\octrl{1}&\qw& \rstick{$  \ket{  n_i  }_{x}  $} & \\  %
 \lstick{$  \ket{  0  }  $} &\targ{}&\gate[2]{A'}&\qw&&&\qw&\qw&&&\gate[2]{A'}&\targ{}&\qw& \rstick{$  \ket{  0  }  $} & \\  %
 \lstick{$  \ket{  0  }  \ket{  n_{\ell}  }_{x}  $} & \qw  \qwbundle{N+2} && \qw  \qwbundle{N+2} && \gate{X^{\otimes  N+2}} &\gate{C^{N+1}(Z)}& \gate{X^{\otimes  N+2}} && \qw  \qwbundle{N+2} &&\qw& \qw  \qwbundle{N+2} & \rstick{$  \ket{  0  }  \ket{  n_{\ell}  }_{x}  $} & \\  %
 \lstick{$  \ket{  0  }  $} &\qw&\qw&\qw&\qw&\gate{H}&\ctrl{-1}&\gate{H}&\qw&\qw&\qw&\qw&\qw& \rstick{$  \ket{  F  }  $} & \\  %
 \end{quantikz}  
  \caption{\label{fig:aglCircuit-qq}
    A routine for checking the Abelian Gauss law along a link joining two quark sites.
    The registers here are specific to a 1D link, with the constraint $[ n_\ell + n_o (1-n_i) ]_x = [ n_\ell + n_i (1-n_o ) ]_{x+1}$.
  }
  \end{center}
\end{figure*}
\section{\label{sec:circuits}Implementing gauge invariance}

Having access to ``Gauss law oracles'' within a simulation opens the door to designing error detection and mitigation protocols that help to preserve gauge invariance.
Such oracles seems especially crucial to making noisy simulations robust against unphysical errors.
Thanks to the diagonalization of the gauge constraints, the space of allowed states is akin to a computational ``code space,'' and validating wavefunctions is likened to measurement of the appropriate ``stabilizers'';
Gauss's law itself defines a code space within a larger Hilbert space that is stabilized by any charge-conserving Hamiltonian.
The associated parity can be thought of as $+1$ for the Abelian-Gauss-law-satisfying subspace and $-1$ for the rest of the states.
Detection of a large class of bit-flip errors can therefore be achieved without any additional encoding of the qubits.

In terms of quantum numbers in the LSH basis, the constraints on states take the following forms:
\begin{itemize}
  \item For $d=1$
\begin{equation}
  \label{eq:1d_qq_agl}[ n_\ell + n_o (1-n_i) ]_x = [ n_\ell + n_i (1-n_o ) ]_{x+1} \ .
\end{equation}
  \item For $d=2$
\begin{subequations}
\begin{align}
  \label{eq:2d_gg_agl} n_j(x^{\prime}) &= n_{\bar j}(\overline{x+e_j}^{\,\prime})  \qquad (j=1,2), \\
  \label{eq:2d_3i_agl} n_3 (x^\prime)  &=  n_l(x) + n_i(x) [1-n_o(x)] , \\
  n_{\bar{3}} (\bar{x}^{\prime}) &= n_l(x) + n_o(x) [1-n_i(x)] ,
\end{align}
\end{subequations}
where $n_1=l_{31}+l_{12}$, $n_2=l_{12}+l_{23}$, and $n_3=l_{23}+l_{31}$ are the eigenvalues from (\ref{eq:gluonicFluxNumbers}).
 \item For $d=3$
\begin{subequations}
\begin{align}
  n_i(x^{\prime}) &= n_{\bar j}(\overline{x+e_j}^{\,\prime})  \qquad (j=1,2), \\
  n_3(x '') &= n_{\bar{3}}(\overline{x+e_3}'') \ , \\
  n_5(x ') &= n_{\bar{5}}(\bar{x}'') \ , \\
  n_6(x '') &= n_{\bar{6}}(\bar{x}') \ , \\
  n_4(x '') &= n_i(x) [1-n_o(x)] \ , \\
  n_{\bar{4}}(\bar{x} '') &= n_o(x) [1-n_i(x)] \ .
\end{align}
\end{subequations}
\end{itemize}

In Fig.~\ref{fig:aglCircuit-qq} we give a quantum circuit for checking the Abelian Gauss law along a ($d=1$) $qq$ link.
The non-standard circuit notations used are the bit-adder gates $A'$, which add one bit to an $M$-bit integer via 
\begin{align*}
    \ket{ y } \ket{0}^{\otimes M} \ket{ c_0 }   {\rightarrow} \ket{y+c_0} \ket{ c_{M-1} c_{M-2} \ldots c_1 } \ket{c_0}  
\end{align*}
(see Fig.~\ref{fig:adder-gates}), and the string of controlled \textsc{NOT}si (\CNOT s), used to compute the \emph{bitwise sum} of two integers, i.e.,
\begin{align*}
  & \ket{x_{r-1}} \ket{x_{r-2}} \cdots \ket{x_0} \otimes \ket{y_{r-1}} \ket{y_{r-2}} \cdots \ket{y_0}\\
  \rightarrow & \ket{x_{r-1}} \ket{x_{r-2}} \cdots \ket{x_0} \otimes \\
  & \qquad \otimes \ket{y_{r-1} \oplus x_{r-1}} \ket{y_{r-2} \oplus x_{r-2}} \cdots \ket{y_0 \oplus x_0} \ .
\end{align*}
\begin{figure*}
\begin{center}
  \begin{tikzpicture}
    \coordinate (A) at ( 0, 0);
    \coordinate (2dOrigin) at ( 4.25, 0);
    \coordinate (C) at ( 9, 0);
    \coordinate (D) at ( 4.5, -3.75);

    \node at (A) {
    \begin{quantikz}[row  sep=0.3cm]
     \\
     \\
     & \lstick{ $ \ket{c_0} $ } & \gate[3]{A'} & \qw & \rstick{ $ \ket{c_0} $ } \\
     & \lstick{ $ \ket{0} \ket{y} $ } &  & \qw & \rstick{ $ \ket{y+c_0} $ } \\ 
     & \lstick{ $ \ket{0}^{\otimes (M-1)} $ } &  & \qw & \rstick{ $ \ket{ c_{M-1} c_{M-2} \ldots c_1 } $ } \\ 
     \\ 
      & \lstick{ $ \ket{ y } \ket{0}^{\otimes M} \ket{ c_0 }$ } & {\rightarrow} &  \rstick{ $ \ket{y+c_0} \ket{ c_{M-1} c_{M-2} \ldots c_1 } \ket{c_0} $ } \\ 
    \end{quantikz}
    };

    \draw (2dOrigin) node {$\quad=\quad$};
    \draw ($(2dOrigin) + ( 0, -2.75)$) ;

    \node at (C) {
    \begin{quantikz}[row  sep=0.35cm]  
    \lstick{$  \ket{  c_0  }  $} &\ctrl{1}&\qw&\qw&\ctrl{1}&\qw& \rstick{$  \ket{  c_0  }  $} & \\  %
    \lstick{$  \ket{  y_0  }  $} &\ctrl{1}&\qw&\qw&\targ{}&\qw& \rstick{$  \ket{  s_0  =  y_0  \oplus  c_0  }  $} & \\  %
    \lstick{$  \ket{  0  }  $} &\targ{}&\ctrl{1}&\qw&\ctrl{1}&\qw& \rstick{$  \ket{  c_1  =  y_0  c_0  }  $} & \\  %
    \lstick{$  \ket{  y_1  }  $} &\qw&\ctrl{1}&\qw&\targ{}&\qw& \rstick{$  \ket{  s_1  =  y_1  \oplus  c_1  }  $} & \\  %
    \lstick{$  \ket{  0  }  $} &\qw&\targ{}&\ctrl{1}&\ctrl{1}&\qw& \rstick{$  \ket{  c_2  =  y_1  y_0  c_0}  $} & \\  %
    \lstick{$  \ket{  y_2  }  $} &\qw&\qw&\ctrl{1}&\targ{}&\qw& \rstick{$  \ket{  s_2  =  y_2  \oplus  c_2  }  $} & \\  %
    \lstick{$  \ket{  0  }  $} &\qw&\qw&\targ{}&\qw&\qw& \rstick{$  \ket{  c_3  =  y_2  y_1  y_0  c_0  =  s_3  }  $} & \\  %
    \end{quantikz}  
    };

    \node at (D) {
    \begin{quantikz}[transparent]  
    \lstick{$  \ket{y}  $} & \qw  \qwbundle{M} &\gate[2]{A}& \qw  \qwbundle{M} &\qw& \rstick{$  \ket{y}  $} & \\  %
    \lstick{$  \ket{0}  \ket{z}  $} & \qw  \qwbundle{M+1} && \qw  \qwbundle{M+1} &\qw& \rstick{$  \ket{z+y}  $} & \\  %
    \end{quantikz}  
    };
  \end{tikzpicture}
  \caption{\label{fig:adder-gates}
    Top: A reduced adder $A'$ for adding one bit to an $M$-bit integer (shown for $M=3$).
    Here $c_0$ is added to the $M$-bit integer $y$, with the $(M+1)$-bit result $s=y+c_0$.
    This uses $M$ ancillae, $M$ Toffoli gates, and $M$ \CNOT s .
    Bottom: A generic adder circuit $A$ for in-place addition of two $M$-bit integers, $(y,z)\rightarrow(y,z+y)$.
    One ancilla is introduced to express the $(M+1)$-bit output.
  }
\end{center}
\end{figure*}
The constraint (\ref{eq:1d_qq_agl}) is equivalent to saying that the $(N+2)$-bit integers $n_L(x) = [ n_\ell + n_o (1-n_i) ]_x$ and $n_R(x+1) = [ n_\ell + n_i (1-n_o ) ]_{x+1}$ are identical bit by bit.
So the constraint is checked by
(i) computing the sums $ n_L(x)$ and $n_R(x+1)$,
(ii) computing the bit-wise sum of $n_L(x)$ and $n_R(x+1)$,
and (iii) flagging the query qubit if and only if that bit-wise sum is all zeros.
The subsequent gates just uncompute the nonquery registers back to their original configurations.
The query output $F=1$ ($F=0$) if the link does (does not) satisfy the Abelian Gauss law.

\begin{figure*}
  \begin{center}
 \begin{quantikz}[row  sep=0.35cm]  
 \lstick{$  \ket{  \ell_{23}  }  $} & \qw  \qwbundle{N} &\gate[2]{A}&\qw&\qw&\qw& \qw  \qwbundle{N} &\qw&\qw&\qw&\gate[2]{A^\dagger}&\qw& \qw  \qwbundle{N} & \rstick{$  \ket{  \ell_{23}  }  $} & \\  %
 \lstick{$  \ket{0}  \ket{  \ell_{31}  }  $} & \qw  \qwbundle{N+1} && \qw  \qwbundle{N+1} & \gate[5,nwires={2,3,4},disable  auto  height]{\begin{tikzpicture}[scale=0.6]  \coordinate (A) at ( 0.25, -0.4); \coordinate (B) at ( 0.5, -0.8); \coordinate (C) at ( 0.75, -1.2); \coordinate (D) at ( 1.0, -1.6); \draw ($( 0, 0) + ( 0, 0)$) -- ($ ( 1.25, 0) + ( 0, 0)$); \foreach \dy in { -0.4, -0.8, -1.6}{   \draw ($( 0, 3) + ( 0, \dy)$) -- ($ ( 1.25, 3) + ( 0, \dy)$);   \draw ($( 0, 0) + ( 0, \dy)$) -- ($ ( 1.25, 0) + ( 0, \dy)$); }; \filldraw [fill=black, draw=black] ($ (A) + ( 0, 3) $)  circle [radius=0.0625]; \filldraw [fill=black, draw=black] ($ (B) + ( 0, 3) $)  circle [radius=0.0625]; \filldraw [fill=black, draw=black] ($ (D) + ( 0, 3) $)  circle [radius=0.0625]; \draw (A) circle [radius=0.125]; \draw ($(A) + ( -0.125, 0)$) -- ($(A) + ( +0.125, 0)$); \draw ($(A) + ( 0, -0.125)$) -- ($(A) + ( 0, +0.125)$); \draw (B) circle [radius=0.125]; \draw ($(B) + ( -0.125, 0)$) -- ($(B) + ( +0.125, 0)$); \draw ($(B) + ( 0, -0.125)$) -- ($(B) + ( 0, +0.125)$); \draw (D) circle [radius=0.125]; \draw ($(D) + ( -0.125, 0)$) -- ($(D) + ( +0.125, 0)$); \draw ($(D) + ( 0, -0.125)$) -- ($(D) + ( 0, +0.125)$); \foreach \y in { 3, 0} {   \draw ($ (C) + ( 0, \y) $) node {$\cdot$};   \draw ($ (C) + ( 0, \y) + ( -0.0625, 0.1) $) node {$\cdot$};   \draw ($ (C) + ( 0, \y)  + ( 0.0625, -0.1)$) node {$\cdot$}; }; \draw (A) -- ($ (A) + ( 0, 3) $); \draw (B) -- ($ (B) + ( 0, 3) $); \draw (D) -- ($ (D) + ( 0, 3) $);   \end{tikzpicture}  }   &\qw& \qw  \qwbundle{N+1} &\qw& \gate[5,nwires={2,3,4},disable  auto  height]{\begin{tikzpicture}[scale=0.6]  \coordinate (A) at ( 0.25, -0.4); \coordinate (B) at ( 0.5, -0.8); \coordinate (C) at ( 0.75, -1.2); \coordinate (D) at ( 1.0, -1.6); \draw ($( 0, 0) + ( 0, 0)$) -- ($ ( 1.25, 0) + ( 0, 0)$); \foreach \dy in { -0.4, -0.8, -1.6}{   \draw ($( 0, 3) + ( 0, \dy)$) -- ($ ( 1.25, 3) + ( 0, \dy)$);   \draw ($( 0, 0) + ( 0, \dy)$) -- ($ ( 1.25, 0) + ( 0, \dy)$); }; \filldraw [fill=black, draw=black] ($ (A) + ( 0, 3) $)  circle [radius=0.0625]; \filldraw [fill=black, draw=black] ($ (B) + ( 0, 3) $)  circle [radius=0.0625]; \filldraw [fill=black, draw=black] ($ (D) + ( 0, 3) $)  circle [radius=0.0625]; \draw (A) circle [radius=0.125]; \draw ($(A) + ( -0.125, 0)$) -- ($(A) + ( +0.125, 0)$); \draw ($(A) + ( 0, -0.125)$) -- ($(A) + ( 0, +0.125)$); \draw (B) circle [radius=0.125]; \draw ($(B) + ( -0.125, 0)$) -- ($(B) + ( +0.125, 0)$); \draw ($(B) + ( 0, -0.125)$) -- ($(B) + ( 0, +0.125)$); \draw (D) circle [radius=0.125]; \draw ($(D) + ( -0.125, 0)$) -- ($(D) + ( +0.125, 0)$); \draw ($(D) + ( 0, -0.125)$) -- ($(D) + ( 0, +0.125)$); \foreach \y in { 3, 0} {   \draw ($ (C) + ( 0, \y) $) node {$\cdot$};   \draw ($ (C) + ( 0, \y) + ( -0.0625, 0.1) $) node {$\cdot$};   \draw ($ (C) + ( 0, \y)  + ( 0.0625, -0.1)$) node {$\cdot$}; }; \draw (A) -- ($ (A) + ( 0, 3) $); \draw (B) -- ($ (B) + ( 0, 3) $); \draw (D) -- ($ (D) + ( 0, 3) $);   \end{tikzpicture}  }   & \qw  \qwbundle{N+1} &&\qw& \qw  \qwbundle{N+1} & \rstick{$  \ket{0}  \ket{  \ell_{31}  }  $} & \\  %
 \lstick{$  \ket{  n_i  }  $} &\ctrl{1}&\qw&\qw&&&\qw&\qw&&&\qw&\ctrl{1}&\qw& \rstick{$  \ket{  n_i  }  $} & \\  %
 \lstick{$  \ket{  n_o  }  $} &\octrl{1}&\qw&\qw&&&\qw&\qw&&&\qw&\octrl{1}&\qw& \rstick{$  \ket{  n_o  }  $} & \\  %
 \lstick{$  \ket{  0  }  $} &\targ{}&\gate[2]{A'}&\qw&&&\qw&\qw&&&\gate[2]{A'^\dagger}&\targ{}&\qw& \rstick{$  \ket{  0  }  $} & \\  %
 \lstick{$  \ket{  0  }  \ket{  n_{\ell}  }  $} & \qw  \qwbundle{N+2} && \qw  \qwbundle{N+2} && \gate{X^{\otimes  N+2}} &\gate{C^{N+1}(Z)}& \gate{X^{\otimes  N+2}} && \qw  \qwbundle{N+2} &&\qw& \qw  \qwbundle{N+2} & \rstick{$  \ket{0}  \ket{  n_{\ell}  }  $} & \\  %
 \lstick{$  \ket{  0  }  $} &\qw&\qw&\qw&\qw&\gate{H}&\ctrl{-1}&\gate{H}&\qw&\qw&\qw&\qw&\qw& \rstick{$  \ket{  F  }  $} & \\  %
 \end{quantikz}  
  \caption{\label{fig:aglCircuit-qg}
    A routine for checking the Abelian Gauss law along a link joining a quark site and a gluonic site.
    The registers here are specific to a 3-$i$ (virtual) link in two dimensions, with the constraint $\ell_{23}+\ell_{31} = n_\ell + n_i (1-n_o)$.
    The adders $A$ and string of \CNOT s are detailed in Fig.~\ref{fig:adder-gates} and in the text, respectively.
  }
  \end{center}
\end{figure*}
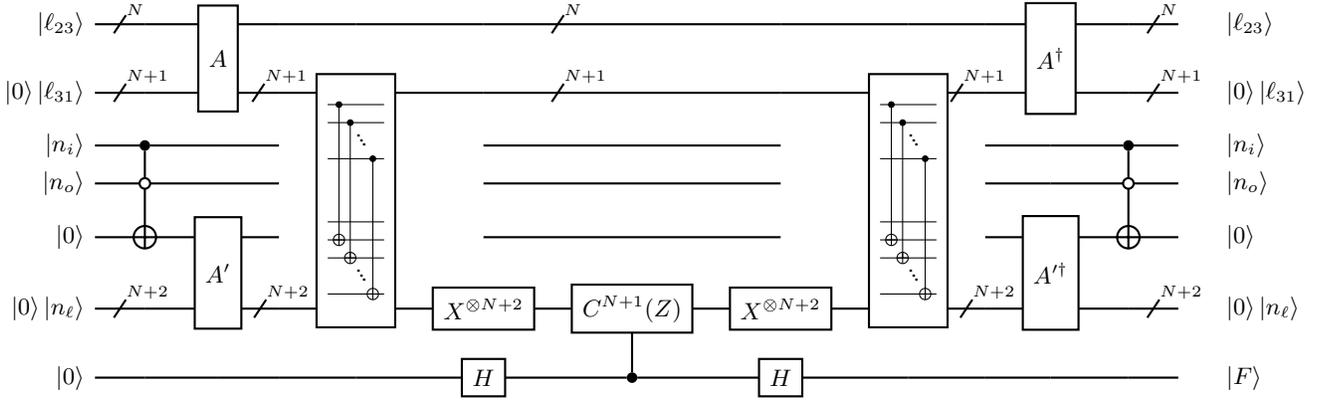
In Fig.~\ref{fig:aglCircuit-qg} we give a quantum circuit for checking the Abelian Gauss law along a $qg$ link in $d\geq2$.
The additional non-standard circuit notation introduced is the $N$-bit addition gate $A$, detailed in Fig.~\ref{fig:adder-gates}.
The input and output registers specifically consider (\ref{eq:2d_3i_agl}), i.e., the Abelian Gauss law across a $3-i$ link of the 2D lattice.
This constraint is checked by
(i) computing the sums $n_3=\ell_{23}+\ell_{31}$ and $n_R = n_\ell + n_i (1-n_o)$,
(ii) computing the bit-wise sum of $n_3$ and $n_{R}$,
and (iii) flagging the query qubit if and only if that bitwise sum is all zeros.

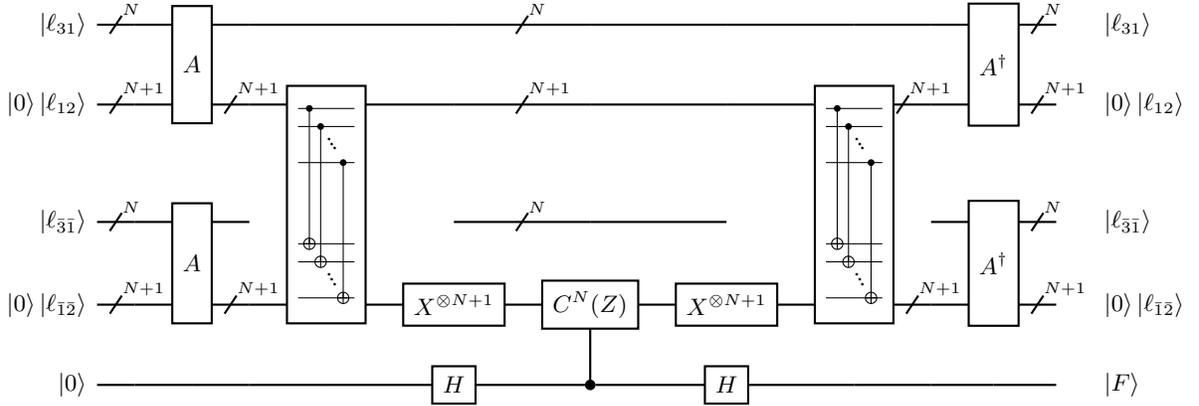
\begin{figure*}
  \begin{center}
 \begin{quantikz}  
 \lstick{$  \ket{  \ell_{31}  }  $} & \qw  \qwbundle{N} &\gate[2]{A}&\qw&\qw&\qw& \qw  \qwbundle{N} &\qw&\qw&\qw&\gate[2]{A^\dagger}& \qw  \qwbundle{N} & \rstick{$  \ket{  \ell_{31}  }  $} & \\  %
 \lstick{$  \ket{0}  \ket{  \ell_{12}  }  $} & \qw  \qwbundle{N+1} && \qw  \qwbundle{N+1} & \gate[4,nwires={2,3},disable  auto  height]{\begin{tikzpicture}[scale=0.6]  \coordinate (A) at ( 0.25, -0.4); \coordinate (B) at ( 0.5, -0.8); \coordinate (C) at ( 0.75, -1.2); \coordinate (D) at ( 1.0, -1.6); \foreach \dy in { -0.4, -0.8, -1.6}{   \draw ($( 0, 3) + ( 0, \dy)$) -- ($ ( 1.25, 3) + ( 0, \dy)$);   \draw ($( 0, 0) + ( 0, \dy)$) -- ($ ( 1.25, 0) + ( 0, \dy)$); }; \filldraw [fill=black, draw=black] ($ (A) + ( 0, 3) $)  circle [radius=0.0625]; \filldraw [fill=black, draw=black] ($ (B) + ( 0, 3) $)  circle [radius=0.0625]; \filldraw [fill=black, draw=black] ($ (D) + ( 0, 3) $)  circle [radius=0.0625]; \draw (A) circle [radius=0.125]; \draw ($(A) + ( -0.125, 0)$) -- ($(A) + ( +0.125, 0)$); \draw ($(A) + ( 0, -0.125)$) -- ($(A) + ( 0, +0.125)$); \draw (B) circle [radius=0.125]; \draw ($(B) + ( -0.125, 0)$) -- ($(B) + ( +0.125, 0)$); \draw ($(B) + ( 0, -0.125)$) -- ($(B) + ( 0, +0.125)$); \draw (D) circle [radius=0.125]; \draw ($(D) + ( -0.125, 0)$) -- ($(D) + ( +0.125, 0)$); \draw ($(D) + ( 0, -0.125)$) -- ($(D) + ( 0, +0.125)$); \foreach \y in { 3, 0} {   \draw ($ (C) + ( 0, \y) $) node {$\cdot$};   \draw ($ (C) + ( 0, \y) + ( -0.0625, 0.1) $) node {$\cdot$};   \draw ($ (C) + ( 0, \y)  + ( 0.0625, -0.1)$) node {$\cdot$}; }; \draw (A) -- ($ (A) + ( 0, 3) $); \draw (B) -- ($ (B) + ( 0, 3) $); \draw (D) -- ($ (D) + ( 0, 3) $);  \end{tikzpicture}  }   &\qw& \qw  \qwbundle{N+1} &\qw& \gate[4,nwires={2,3},disable  auto  height]{\begin{tikzpicture}[scale=0.6]  \coordinate (A) at ( 0.25, -0.4); \coordinate (B) at ( 0.5, -0.8); \coordinate (C) at ( 0.75, -1.2); \coordinate (D) at ( 1.0, -1.6); \foreach \dy in { -0.4, -0.8, -1.6}{   \draw ($( 0, 3) + ( 0, \dy)$) -- ($ ( 1.25, 3) + ( 0, \dy)$);   \draw ($( 0, 0) + ( 0, \dy)$) -- ($ ( 1.25, 0) + ( 0, \dy)$); }; \filldraw [fill=black, draw=black] ($ (A) + ( 0, 3) $)  circle [radius=0.0625]; \filldraw [fill=black, draw=black] ($ (B) + ( 0, 3) $)  circle [radius=0.0625]; \filldraw [fill=black, draw=black] ($ (D) + ( 0, 3) $)  circle [radius=0.0625]; \draw (A) circle [radius=0.125]; \draw ($(A) + ( -0.125, 0)$) -- ($(A) + ( +0.125, 0)$); \draw ($(A) + ( 0, -0.125)$) -- ($(A) + ( 0, +0.125)$); \draw (B) circle [radius=0.125]; \draw ($(B) + ( -0.125, 0)$) -- ($(B) + ( +0.125, 0)$); \draw ($(B) + ( 0, -0.125)$) -- ($(B) + ( 0, +0.125)$); \draw (D) circle [radius=0.125]; \draw ($(D) + ( -0.125, 0)$) -- ($(D) + ( +0.125, 0)$); \draw ($(D) + ( 0, -0.125)$) -- ($(D) + ( 0, +0.125)$); \foreach \y in { 3, 0} {   \draw ($ (C) + ( 0, \y) $) node {$\cdot$};   \draw ($ (C) + ( 0, \y) + ( -0.0625, 0.1) $) node {$\cdot$};   \draw ($ (C) + ( 0, \y)  + ( 0.0625, -0.1)$) node {$\cdot$}; }; \draw (A) -- ($ (A) + ( 0, 3) $); \draw (B) -- ($ (B) + ( 0, 3) $); \draw (D) -- ($ (D) + ( 0, 3) $);  \end{tikzpicture}  }   & \qwbundle{N+1}  \qw && \qw  \qwbundle{N+1} & \rstick{$  \ket{0}  \ket{  \ell_{12}  }  $} & \\  %
 \\  
 \lstick{$  \ket{  \ell_{\bar{3}\bar{1}}  }  $} & \qw  \qwbundle{N} &\gate[2]{A}&\qw&&& \qw  \qwbundle{N} &\qw&&&\gate[2]{A^\dagger}& \qw  \qwbundle{N} & \rstick{$  \ket{  \ell_{\bar{3}\bar{1}}  }  $} & \\  %
 \lstick{$  \ket{0}  \ket{  \ell_{\bar{1}\bar{2}}  }  $} & \qw  \qwbundle{N+1} && \qw  \qwbundle{N+1} && \gate{X^{\otimes  N+1}} &\gate{C^N(Z)}& \gate{X^{\otimes  N+1}} && \qwbundle{N+1}  \qw && \qw  \qwbundle{N+1} & \rstick{$  \ket{0}  \ket{  \ell_{\bar{1}\bar{2}}  }  $} & \\  %
 \lstick{$  \ket{  0  }  $} &\qw&\qw&\qw&\qw&\gate{H}&\ctrl{-1}&\gate{H}&\qw&\qw&\qw&\qw& \rstick{$  \ket{  F  }  $} & \\  %
 \end{quantikz}  
  \caption{\label{fig:aglCircuit-gg}
    A routine for checking the Abelian Gauss law along a link joining two gluonic sites.
    The registers here are specific to a 1-$\bar{1}$ (physical direction) link in 2D space, with the constraint $\ell_{23}(x)+\ell_{31}(x) = \ell_{\bar{2}\bar{3}}(x+e_1)+\ell_{\bar{3}\bar{1}}(x+e_1)$.
    The adder circuits $A$ are detailed in Fig.~\ref{fig:adder-gates}, and the \CNOT  string is detailed in the text.
  }
  \end{center}
\end{figure*}
In Fig.~\ref{fig:aglCircuit-gg}, we give a quantum circuit for checking the Abelian Gauss law along a $gg$ link in $d\geq2$.
The input and output registers in Fig.~\ref{fig:aglCircuit-gg} specifically apply to (\ref{eq:2d_gg_agl}), namely, the Abelian Gauss law across a $1-\bar{1}$ $gg$ link of the 2D lattice.
The circuit
(i) separately computes $n_1=\ell_{31}(x)+\ell_{12}(x)$ and $n_{\bar{1}}=\ell_{\bar{3}\bar{1}}(x+e_1)+\ell_{\bar{1}\bar{2}}(x+e_1)$,
(ii) computes the bitwise sum of $n_1$ and $n_{\bar{1}}$,
and (iii) flags the query output qubit if and only if that bitwise sum is all zeros.

In Table \ref{tab:circuitCosts}, we give resource costs for the oracles using simple algorithms from the literature.
\begin{table*}
  \centering
  \begin{tabular}{c | c | c | c | c | c | c | c | c | c | c}
    {} & \multicolumn{3}{c|}{Explicit components} & \multicolumn{3}{c|}{$A$ gate adders} & \multicolumn{3}{c|}{$A'$ gate adder} & \multicolumn{1}{c|}{Multicontrolled $Z$ \cite{barenco.bennett.eaElementaryGates95}} \\
    {} & \multicolumn{3}{c|}{} & \multicolumn{3}{c|}{(with $A$ using ripple-carry \cite{cuccaro.draper.eaNewQuantum04})} & \multicolumn{3}{c|}{($A'$ as in Fig.~\ref{fig:adder-gates})} & \multicolumn{1}{c|}{(using LSH workspace)} \\
    {Circuit} & Anc. & \CNOT & $C^2(X)$ cong. & Anc. & \CNOT & $C^2(X)$ cong. & Anc. & \CNOT & $C^2(X)$ cong. & exact $C^2(X)$ \\
    \hline
    $qq$ & 4 & $2(N+2)$ & $4$ & - & - & - & $2(N+1)$ & $4(N+1)$ & $4(N+1)$ & $4N$ \\
    $qg$ & 3 & $2(N+1)$ & $2$ & $1$ & $2(5N-3)$ & $2(2N-1)$ & $N+1$ & $2(N+1)$ & $2(N+1)$ & $4N$ \\
    $gg$ & 2 & $2(N+1)$ & $0$ & $2$ & $4(5N-3)$ & $4(2N-1)$ & - & - & - & $4(N-1)$ \\
  \end{tabular}
  \caption{\label{tab:circuitCosts}
  Resource counts for implementing the oracles using simple subroutines from the literature.
  $N\geq3$ refers to the gluonic loop number register size;
  the quark site loop number register size is $N+1$.
  The counted resources are numbers of ancillary qubits (Anc.), \CNOT  gates, and Toffoli-congruent [$C^2(X)$ cong.] or exact Toffoli [$C^2(X)$] gates.
}
\end{table*}
The main measure of complexity considered is the number of Toffoli gates, broken down into ``Toffoli-congruent'' gates and exact Toffolis.
By Toffoli-congruent gates we mean three-qubit gates that map computational basis states like a Toffoli but with possible phase shifts.
The freedom of phase shifts means Toffoli-congruent gates can be implemented more efficiently than exact Toffolis \cite{barenco.bennett.eaElementaryGates95}.
Toffoli-congruent gates are acceptable in the oracles' adders since the undesired phases get removed during the uncomputation cycle.
However, these phases cannot be introduced in the multi-controlled $Z$ operation.

To give complete gate counts we had to choose algorithms for the subroutines.
We have already chosen to implement the bit adders $A'$ in the tailored way given earlier.
For the adder gates $A$, we considered the ripple-carry addition algorithm (in place and with no incoming carry bit) from \cite{cuccaro.draper.eaNewQuantum04}.
And for the multicontrolled $Z$ operations, we apply Lemma 7.2 of \cite{barenco.bennett.eaElementaryGates95} by using some of the nonparticipating LSH registers as work space;
in the $gg$ circuit the $N$-qubit work space could be the $\ket{\ell_{\bar{3}\bar{1}}}$ register, while in the $qg$ circuit the $(N+1)$-qubit workspace could be the $\ket{0}\ket{\ell_{31}}$ register.
In applying the algorithms of Refs.~\cite{barenco.bennett.eaElementaryGates95,cuccaro.draper.eaNewQuantum04}, $N\geq3$ is assumed.

By taking advantage of Toffoli congruence and using Corollary 6.2 of \cite{barenco.bennett.eaElementaryGates95} for the exact Toffolis, we can consolidate the multiqubit gate counts into equivalent \CNOT  counts.
These are reported in Table \ref{tab:CNOTCounts}.
\begin{table}
  \centering
  \begin{tabular}{ c | l }
    Circuit & Equivalent \CNOT s \\
    \hline
    $qq$ & $42N+32$ \\
    $qg$ & $56N+4$ \\
    $gg$ & $70N-46$
  \end{tabular}
  \caption{\label{tab:CNOTCounts}
    \CNOT  counts for the circuits assuming $N\geq3$, the use of Toffoli-congruent gates, and Corollary 6.2 of \cite{barenco.bennett.eaElementaryGates95}.
}
\end{table}

\section{\label{sec:discussion}Discussion}

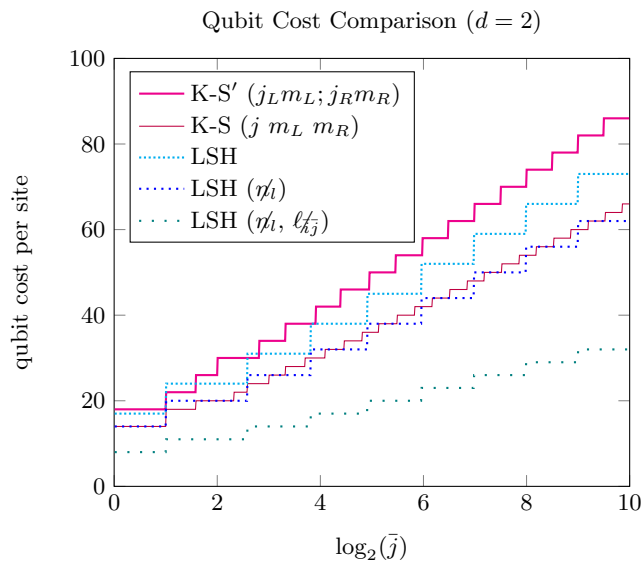
\begin{figure}
  \begin{center}
  \begin{tikzpicture}
    \begin{axis}[xmin=0, xmax=10, domain=0.01:9.99,
        ymin=0, ymax=100,
        xlabel=$\log_2(\bar{j})$, ylabel={qubit cost per site},
        title = {Qubit Cost Comparison ($d=2$)},
        legend pos = north west,
        legend cell align={left}
    ]
      \addplot+[no marks, solid, thick, magenta, samples=1000] ({x}, {
      2 + 4 * ceil( ln( ( 1 + ceil(2^x) ) * (1 + 2 * ceil(2^x) )) / ln(2) )
      });
      \addlegendentry{K-S$'$ ($j_L m_L ; j_R m_R$)};

      \addplot+[no marks, solid, thin, purple, samples=1000] ({x}, {
      2 + 2 * ceil( ln( 8/3 * (ceil(2^x)+1/2) * (ceil(2^x)+3/4) * (ceil(2^x)+1)) / ln(2))
      });
      \addlegendentry{K-S ($j ~ m_L ~ m_R $)};

      \addplot+[no marks, densely dotted, thick, cyan, samples=1000] ({x}, {
      3 + 7 * ceil( ln( 1 + ceil(2^x) ) / ln(2) )
      });
      \addlegendentry{LSH};

      \addplot+[no marks, dotted, thick, blue, samples=1000] ({x}, {
      2 + 6 * ceil( ln( 1 + ceil(2^x) ) / ln(2) )
      });
      \addlegendentry{LSH ($\centernot{n_l}$)};

      \addplot+[no marks, loosely dotted, thick, teal, samples=1000] ({x}, {
      2 + 3 * ceil( ln( 1 + ceil(2^x) ) / ln(2) )
      });
      \addlegendentry{LSH ($\centernot{n_l}$, $\centernot{\ell_{\bar{i}\bar{j}}}$)};
    \end{axis}
  \end{tikzpicture}
  \caption{
    \label{fig:qubitCosts2d}
    The qubit costs per (unsplit) 2D lattice site as a function of the Casimir cutoff $\bar{j}$.
    The bases considered are the conventional Kogut-Susskind (K-S, thin solid line), loop-string-hadron (LSH), and modifications of these as described in the text.
}
  \end{center}
\end{figure}

\begin{figure}
  \begin{center}
  \begin{tikzpicture}
    \begin{axis}[xmin=0, xmax=10, domain=0.01:9.99,
        ymin=0, ymax=150,
        xlabel=$\log_2(\bar{j})$, ylabel={qubit cost per site},
        title = {Qubit Cost Comparison ($d=3$)},
        legend pos = north west,
        legend cell align={left}
    ]
      \addplot+[no marks, solid, thin, magenta, samples=1000] ({x}, {
      2 + 6 * ceil( ln( ( 1 + ceil(2^x) ) * (1 + 2 * ceil(2^x) )) / ln(2) )
      });
      \addlegendentry{K-S$'$ ($j_L m_L ; j_R m_R$)};

      \addplot+[no marks, solid, thick, purple, samples=1000] ({x}, {
      2 + 3 * ceil( ln( 8/3 * (ceil(2^x)+1/2) * (ceil(2^x)+3/4) * (ceil(2^x)+1)) / ln(2))
      });
      \addlegendentry{K-S ($j ~ m_L ~ m_R $)};

      \addplot+[no marks, densely dotted, thick, cyan, samples=1000] ({x}, {
      3 + 13 * ceil( ln( 1 + ceil(2^x) ) / ln(2) )
      });
      \addlegendentry{LSH};

      \addplot+[no marks, dotted, thick, blue, samples=1000] ({x}, {
      2 + 12 * ceil( ln( 1 + ceil(2^x) ) / ln(2) )
      });
      \addlegendentry{LSH ($\centernot{n_l}$)};

      \addplot+[no marks, loosely dotted, thick, teal, samples=1000] ({x}, {
      2 + 6 * ceil( ln( 1 + ceil(2^x) ) / ln(2) )
      });
      \addlegendentry{LSH ($\centernot{n_l}$, $\centernot{\ell_{\bar{i}\bar{j}}}$)};
    \end{axis}
  \end{tikzpicture}
  \caption{
    \label{fig:qubitCosts3d}
    The qubit costs per (unsplit) 3D lattice site as a function of $\bar{j}$.
    The bases considered are the conventional Kogut-Susskind (K-S, thick solid line), loop-string-hadron (LSH), and modifications of these as described in the text.
}
  \end{center}
\end{figure}
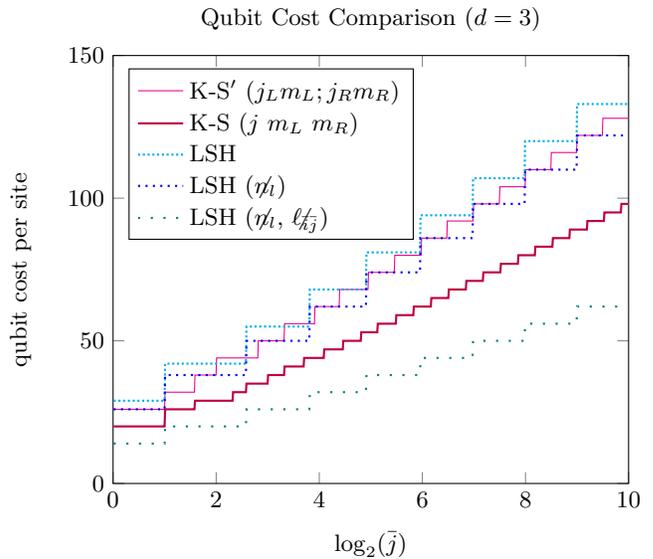

The tradeoff of the LSH approach may be an increase in qubits needed for simulation.
For the truncation level $\bar{j}$ defined above, the total number of logical qubits required for simulating the LSH basis is $$ V \left( \ \left\{ 6(d-1) \lceil \log_2 (\bar{j}+1) \rceil \right\} + \left\{ \lceil \log_2 (\bar{j}+1) \rceil+3 \right\} \ \right) \ ,$$ where $V$ is the Cartesian volume and the curly brackets separate the costs of gluonic sites and quark sites.
The per-site count stems from the $\ell_{ij}$'s at $2(d-1)$ gluonic vertices, the quantum numbers of the quark vertex, and the definition of $N$ from (\ref{eq:lshTruncation}).
For the Kogut-Susskind representation basis as conventionally formulated \cite{zohar.burrelloFormulationLattice15}, the best-case cost is $$ V \left( \ \left\{ d \left\lceil \log_2 [\tfrac{8}{3} (\bar{j}+\tfrac{1}{2}) (\bar{j}+\tfrac{3}{4}) (\bar{j}+1) ] \right\rceil \right\}+ 2 \ \right) \ .$$
In this case the per-site count arises from counting up all irreducible representation states $\ket{j,m,m'}$ on $d$ links and the quark occupancies.
These costs are displayed in Figs.~\ref{fig:qubitCosts2d} and \ref{fig:qubitCosts3d}.
However, as pointed out earlier, the $\ket{j,m,m'}$ basis seems ill suited to digitization, so a better comparison to make could be with a variation of Kogut-Susskind gauge theory in which the link ends are treated separately:
$\ket{j,m,m'} \rightarrow \ket{j,m}\otimes \ket{j',m'}$ with the physical requirement $j=j'$ along links.
In this case, each site would be associated with $2d$ link ends plus the quark doublet, and the logical qubit costs would be
$$ V ( \left\{ 2d \lceil \log_2 [(\bar{j}+1)(2\bar{j}+1)] \rceil \right\} + 2 ) \ . $$

We see from Figs.~\ref{fig:qubitCosts2d} and \ref{fig:qubitCosts3d} that in 2D space with matter, or in 3D space, the LSH basis indeed calls for more qubits to reach a fixed cutoff than does the conventional representation basis (labeled K-S);
the asymptotic logical qubit cost of LSH with one staggered quark flavor is $16.7\%$ higher in $d=2$ and $44.4\%$ higher in $d=3$.
The persistent redundancy of the Hilbert space is apparently the price paid for trading non-Abelian constraints for Abelian ones.
However, it is a common situation in both quantum computation and in gauge theories where expanding a Hilbert space with extra degrees of freedom can yield a more convenient theoretical description.
Digitization of the conventional representation basis could even benefit from a treatment with separate Hilbert spaces at the link ends (labeled K-S$'$), in which case LSH asymptotically saves $12.5\%$ of qubits in $d=2$ and costs only $8.33\%$ more in $d=3$.

Another possibility is to solve the LSH constraints and cut the number of bosonic degrees of freedom in half, but this may increase the range of the interactions between registers and also complicate the interactions in the Hamiltonian.
Additionally, triangle inequalities involving the dynamical quantum numbers would surface, so some unphysical states would still persist.
Although it is not known yet whether this complete reduction has a net advantage, we display the associated qubit counts in Figs.~\ref{fig:qubitCosts2d} and \ref{fig:qubitCosts3d} [indicated by ($\centernot{n_l}$, $\centernot{\ell_{\bar{i}\bar{j}}}$)].

One could instead settle for a smaller reduction, eliminating the quark-site bosonic variables $n_l$, without running into the aforementioned complications.
This slightly reduced LSH variant costs about the same qubits as standard K-S in $d=2$, and the same as or fewer than K-S with separated link ends in $d=3$.
This LSH reduction is similarly shown in Figs.~\ref{fig:qubitCosts2d} and \ref{fig:qubitCosts3d} [indicated by ($\centernot{n_l}$)].

The essential benefit that was exploited above was the trading of non-commuting Gauss law constraints for commuting Abelian Gauss law constraints.
Thus, no change of basis is necessary to check the constraints, and one can immediately proceed with computation.
For this reason we could easily furnish the first explicit algorithms to validate wave functions of a non-Abelian lattice gauge theory.
The algorithms call for common subroutines and are easily adapted to different qubit mappings or even to qudits.
For illustrative purposes we focused on the most straightforward binary integer representation.

\section{\label{sec:conclusion}Conclusion}

The loop-string-hadron formulation as a basis for digital quantum simulation offers important theoretical advantages.
While the exact same physics as Kogut-Susskind gauge theory is being described, this treatment has expedited the process of turning lattice gauge theory formalism into input for quantum algorithm research.
The physicality diagnostics are likely to be useful in digital simulations because non-gauge invariant errors can easily arise from the Trotter approximation to $e^{-it\hat{H}}$ or from quantum noise \cite{strykerOraclesGauss19}.
Previously, Gauss law oracles had only been decomposed for U(1) and Z(N) gauge theories.
Because the LSH framework has cast all constraints into an Abelian form, those techniques could be ported over for use with SU(2).

The LSH framework also endows SU(2) lattice gauge theory with another similarity to U(1) theories:
the plaquette operators in $d\geq 2$ manifestly decompose into a sum of one-sparse ladder operators.
This commonality presents the opportunity for SU(2) simulations to potentially benefit from algorithms designed for U(1) simulations.
A key distinguishing feature of SU(2) to overcome will be that the nonzero matrix elements of the SU(2) ladder operator are functions of the electric flux, whereas the nonzero matrix elements of the U(1) plaquette operator are all identical.

The approach to decomposing physicality oracles presented should generalize to SU(3) just as well.
In the prepotential formulation of SU(3), one obtains two Abelian constraints along each link, so all the benefits of commuting constraints may carry over.
There are still details to be worked out regarding the point splitting procedure and incorporation of matter before circuits can be furnished.

Much remains to be understood about the quantum simulation of lattice gauge theories.
Even for the simplest gauge groups approaches to Gauss's law are still being researched \cite{meuriceDiscreteAspects20}.
New studies are also underway into the consequences of breaking gauge invariance \cite{halimeh.haukeStaircasePrethermalization20,halimeh.haukeOriginStaircase20}.
In this paper we have made the argument that a more physically-motivated basis can be used to enforce all gauge constraints and that there is reason to believe it will help advance simulation protocols.

\section{Acknowledgments}

We thank D.B.~Kaplan, N.~Klco, and M.J.~Savage for helpful discussions and feedback during the course of this work.
I.R.~is supported by the U.S. Department of Energy (DOE), Office of Science, Office of Advanced Scientific Computing Research (ASCR) Quantum Computing Application Teams program, under Fieldwork Proposal No.~ERKJ347, and by the MCFP.
I.R.~was also supported in part by the Thomas L.~and Margo G.~Wyckoff Endowed Faculty Fellowship.
J.R.S.~was supported by DOE Grant No.~DE-FG02-00ER41132 and by the National Science Foundation Graduate Research Fellowship under Grant No.~DGE-1762114.

\bibliography{lshOracles}

\end{document}